\documentclass{elsart}
\usepackage{graphicx}
\usepackage{amssymb}

\begin{document}
\begin{frontmatter}

\title{Scaling of star polymers: high order results}

 \author[label1]{V.\ Schulte-Frohlinde}
 \ead{frohlind@physik.uni-freiburg.de}
 \author[label2,label3]{Yu.\ Holovatch\corauthref{cor1}}
 \ead{hol@icmp.lviv.ua}
 \corauth[cor1]{Corresponding author.}
%  \ead[url]{http://ph.icmp.lviv.ua/~hol/}
\author[label1,label3]{C.\ von Ferber}
\ead{ferber@physik.uni-freiburg.de}
\author[label1]{A.\ Blumen}
\ead{blumen@physik.uni-freiburg.de}
 \address[label1]{Theoretische Polymerphysik, Universit\"at Freiburg,
                D-79104 Freiburg, Germany}
 \address[label2]{Institute for Condensed Matter Physics
 and Ivan Franko National University of Lviv,
                  UA-79011 Lviv, Ukraine}
 \address[label3]{Institut f\"ur Theoretische Physik, Johannes Kepler
 Universit\"at Linz, A-4040 Linz, Austria}

\begin{abstract}
We extend existing renormalization group calculations for the
exponents describing scaling of star polymers and polymer networks
constituted by chains of different species (the so-called
copolymer star exponents). Our four loop results find application
in the description of various phenomena involving self-avoiding
and random walks that interact.
\end{abstract}

\end{frontmatter}

%%%%%%%%%%%%%%%%%%%%%%%%%%%%
%SECTION I
%%%%%%%%%%%%%%%%%%%%%%%%%%%%

\section{Introduction} \label{I}

Star polymers, i.e. structures of polymer chains that are
chemically linked with one end to a common core attain recently
much attention both in theory, experiment, and technical
applications \cite{note1}. On the experimental side this is due to
their effective interaction properties that interpolate between
those of colloidal particles and single chain polymers. For theory
and simulation they present the most simple albeit non-trivial
examples of polymeric networks which, as shown by theory, provide
all information to derive the scaling properties of general
polymer networks.  For this reason the scaling properties of stars
of long flexible polymer chains in a good solvent are of central
interest \cite{Duplantier89,Schaefer92}. The scaling of single
long flexible polymer chains is perfectly described by a model of
self-avoiding walks resulting in universal scaling exponents that
only depend on the space dimension $d$ \cite{polymerbooks}.
However, the corresponding scaling exponents for star polymers
will also depend on the number of arms $f$ constituting the star
\cite{Duplantier89}.  Building a polymer star from chains of
different species one finds an even richer scaling behavior
\cite{Cates87,Ferbercopol}. The partition function of such a
so-called copolymer star consisting of $f_1$ chains of species 1
and $f_2$ chains of species 2 (each of the same size $R$) will
show power law scaling in $R$ according to \cite{Ferbercopol}:
\begin{equation}\label{1}
{\mathcal Z}^{(*f_1f_2)} \sim
(R/\ell)^{\eta_{f_1f_2}-f_1\eta_{20}-f_2\eta_{02}},
\end{equation}
with a family of copolymer star exponents $\eta_{f_1f_2}$ for
$f_1,f_2=0,1,2,\ldots$. In (\ref{1}) $\ell$ is a microscopic scale
and a fugacity factor is omitted. We consider a species A of self-
and mutually avoiding chains (SAW), species B and B$^\prime$ of
both self and mutually transparent chains (RW) and a species C of
self-transparent but mutually avoiding chains (MAW). Copolymer
stars are built from one or more of these species with mutual
avoidance between any two species. Of special interest are the
exponents of (a) homogeneous polymer stars $\eta^S_{f_A}$, (b)
symmetric and unsymmetric copolymer stars
$\eta^G_{f_Bf_{B^\prime}}$ and $\eta^U_{f_Af_B}$, and (c) of stars
of mutually avoiding walks $\eta^{MAW}_{f_C}$. These exponents
naturally arise when describing e.g. the scaling properties of the
effective interaction of star polymers, the correlations of the
vertices of a branched polymeric structure, the diffusion of
particles near an absorbing polymer, and  the transition of double
stranded to single stranded DNA
\cite{note1,Duplantier99,Baiesi02}. Moreover, the star exponents
$\eta_{f_Af_B}$ uniquely define the scaling of copolymer networks
of arbitrary but fixed topology
\cite{Duplantier89,Schaefer92,Ferbercopol}. The above mentioned
facts together with theoretical reasons (e.g. the copolymer star
exponents are related to the composite operators of polymer field
theory and possess multifractal properties \cite{FerberMF} which
seemingly is in contradiction with stability conditions
\cite{Duplantier91}) explain the high interest in their analysis.

Currently, the field theoretical renormalization group (RG)
\cite{RGbooks} is recognized as the most reliable analytic tool to
calculate numerical values of scaling exponents in a systematic
perturbative way. For the universality class of $O(m)$-symmetric
$\phi^4$ theory exponents are known up to high orders of
perturbation theory: to the 5th order in $\varepsilon=4-d$
expansion \cite{Kleinert9193} and to the 6th/7th order when
calculated directly at fixed space dimension $d=3$ \cite{Guida98}.
However, the star exponents are known so far only to the 3rd order
of perturbation theory \cite{Schaefer92,Ferbercopol}. A principal
reason for this gap is that copolymer star exponents were
introduced only recently \cite{Ferbercopol} and that these
exponents have no direct counterparts in the theory of magnets.

Our main message is given by the analytic expressions for the
copolymer star exponents that we obtained by state-of-the-art RG
methods to  $\varepsilon^4$ accuracy in an
$\varepsilon$-expansion. This is merely one order below the best
known $\varepsilon$-expansion for the critical exponents of the
$m$-vector model \cite{Kleinert9193}. When evaluated numerically
\cite{Schulte-2003,Schulte-2003a}  by appropriate resummation
techniques \cite{RGbooks} the expansions for the exponents lead to
results in nice correspondence with the available exact (2D) and
MC data. This paper is organized as follows: the next, 2nd section
introduces the main expressions relating the copolymer star
exponents to polymer field theory, section \ref{III} presents the
$\varepsilon$-expansions for the three cases discussed above:
$\eta^G_{f_Bf_{B^\prime}}$, $\eta^U_{f_Af_B}$ and
$\eta^{MAW}_{f_C}$. In the last section we conclude and discuss
our results presenting a numerical estimate drawn from our
expansions. Details of our calculations along with numerical
estimates of different quantities involving star exponents are the
subject of a separate publication \cite{Schulte-2003}.

%%%%%%%%%%%%%%%%%%%%%%%%%%%%
%SECTION II
%%%%%%%%%%%%%%%%%%%%%%%%%%%%

\section{The renormalization group relations}
\label{II}

Here, we sketch the framework of polymer field theory suited to
describe copolymer stars. The starting point is the Edwards model
\cite{polymerbooks} generalized to describe a set of $f_1+f_2$
polymer chains in solution with self- and mutual- avoiding
interactions $u_{ab}$. Within this model, the configuration of
each chain $a=1,\dots,f_1+f_2$ is given by a path $r^a(s)$ in
$d$-dimensional space, $0 \leq s \leq S_a$, $S_a$ is the so-called
Gaussian surface of the path $r^a(s)$. The Hamiltonian ${\mathcal
H}_{f_1f_2}$ of the model reads \cite{Schaefer92,Ferbercopol}:
\begin{equation}\label{2}
\hspace{-1em}
 \frac{1}{k_{\rm B}T} {\mathcal H}_{f_1f_2}({\bf r}^a,\{S_a\}) =
\sum_{a=1}^{f_1+f_2}\int_0^{S_a}{\rm d}s
  (\frac{{\rm d}{\bf r}^a(s)}{2{\rm d}s} )^2 +
 \frac{1}{6}\sum_{a,b=1}^{f_1+f_2} u_{ab}
  \int{\rm d} {\bf r} \rho_a({\bf r})\rho_b({\bf r}) ,
\end{equation}
with densities $\rho_a({\bf r})=\int_0^{S_a}{\rm d}s \delta({\bf
r}-{\bf r}^a(s))$. In Eq. (\ref{2}), the first term describes the
chain connectivity whereas the second one stands for the excluded
volume interaction. The partition function of a star of $f_1+f_2$
polymer chains ${\mathcal Z}^{(*f_1f_2)}$ is calculated as a
functional integral ($\int {\mathcal D} [ {\bf r}_a ]$) of the
Boltzmann distribution with the Hamiltonian (\ref{2}) over all
possible configurations of the paths $r^a(s)$. A product of
$\delta$-functions serves as a constraint ensuring all polymer
chains to have a common endpoint:
\begin{equation}\label{3}
\hspace{-1em} {\mathcal Z}^{(*f_1f_2)} \{ S_a \} = \int {\mathcal
D} [ {\bf r}_a ] \exp  \{ -\frac{1}{k_{\rm B}T}{\mathcal
H}_{f_1f_2}({\bf r}_a,\{S_a\})\}
 \prod_{a=2}^{f_1+f_2} \delta({\bf r}_a(0) - {\bf r}_1(0)).
\end{equation}

The mapping of the Edwards model to a $\phi^4$ field theory with
one coupling of $O(m)$-symmetry in the limit $m=0$ is well
established \cite{polymerbooks}. The polymer system we are
currently interested in is the so-called ternary polymer solution,
where polymer chains of two species (1 and 2) are present. It is
described by an Edwards model (\ref{2}) with three couplings
\cite{Schaefertern}:
\begin{equation}\label{4}
u_{ab}= \{u_{11}, u_{22},  u_{12}=u_{21}\}.
\end{equation}
The partition function of the copolymer star (\ref{3}) corresponds
in the field theory to a vertex function with the insertion of a
composite operator $\Pi_a \phi_a$ of a special symmetry
\cite{Ferbercopol}. Ultraviolet divergences occur when the bare
vertex functions are evaluated naively \cite{RGbooks}. We apply
the field theoretical RG approach \cite{Ferbercopol} to remove the
divergences and to study the scaling properties of the vertex
functions of copolymer stars. Since the theory is renormalizable
one can collect all divergences in the so-called renormalization
factors $Z$ and define a finite theory with renormalized
parameters  preserving the structure of the original one. The
renormalized couplings $g_{ab}$ and the renormalization factors
$Z$ depend on the parameter $\kappa$ which gives the scale of the
external momenta in the renormalization procedure. This dependence
is expressed by RG functions defined by the following relations:
 \begin{eqnarray}\label{5}
 \kappa \frac{\rm d}{{\rm d} \kappa} {g}_{aa} &=& {\beta}_{aa}
 (g_{aa}) ,  \hspace{3em} a=1,2,
 \\ \label{6}
\kappa \frac{\rm d}{{\rm d} \kappa} {g}_{12} &=& {\beta}_{12}
 (g_{11},g_{22},g_{12}),
 \\ \label{7}
 \kappa \frac{\rm d}{{\rm d} \kappa} \ln Z_{*f_1f_2} &=&
 \eta_{f_1f_2}(g_{11},g_{22},g_{12}).
 \end{eqnarray}
In Eq. (\ref{7}), $Z_{*f_1f_2}$ stands for the $Z$-factor
renormalizing the vertex function of a composite operator, which
corresponds to the copolymer star partition function (\ref{3}). In
the fixed points of the RG transformation, given by a common zero
of the $\beta$-functions (\ref{5}), (\ref{6}) the anomalous
dimensions $\eta_{f_1f_2}$ of the composite operators  define the
set of exponents for copolymer stars. Explicit expressions for the
star exponents will be given in the next section.

%%%%%%%%%%%%%%%%%%%%%%%%%%%%
%SECTION III
%%%%%%%%%%%%%%%%%%%%%%%%%%%%

\section{$\varepsilon^4$-expansions for the exponents}
\label{III}

The zeros of the $\beta$-functions (\ref{5}), (\ref{6}) correspond
to the fixed points of the theory. To clarify their physical
significance we here refer to the three classes of species A, B,
and C introduced above and denote nontrivial values by $g^*_S$,
$g^*_G$, and $g^*_U$. The fixed points of (\ref{5}) are
$\beta_{11}(g^*_S)=0$ for species A, $\beta_{aa}(0)=0$ for species
B and C. The non-trivial fixed points of (\ref{6}) are
$\beta_{12}(0,0, g^*_G)=0$ for the combination of species B,
B$^\prime$ and for the interchain interaction of species C and
finally $\beta_{12}(g^*_S,0, g^*_U)=0$ for the combination of
species A and B \cite{Schaefertern}. Analyzing the fixed point
structure, one is led basically to the following physically
different situations: in the first case we combine species B and
B$^\prime$ i.e. both sets of $f_1$ polymer chains of species B and
of $f_2$ polymer chains of species B$^\prime$ are
non-selfinteracting RWs, but mutually avoid each other (we refer
to this as to the case ``$G$"). In the second case, ``$U$", the
species A selfinteract ($f_1$ SAWs) and the species B do not
($f_2$ RWs). In the third case, ``$MAW$", only species C of $f_C$
mutually avoiding RWs is present. In our calculations we use the
dimensional regularization and minimal subtraction scheme
\cite{RGbooks}. Most of the star graphs could be derived from the
four-point diagrams of the $\phi^4$ theory. At the four loop
level, however, we encountered a number of independent star
graphs, which we evaluated following the R$^*$-operation scheme
\cite{RGbooks}. Details of our calculations will be presented
elsewhere \cite{Schulte-2003}. We do not display the expressions
for the $\beta$-functions here although they are of special
interest e.g. to study the crossover behaviour of the ternary
polymer solution \cite{Schaefertern}. Our results for the
copolymer star exponents (\ref{1}) read:
\begin{eqnarray}\nonumber
\eta^G_{f_1f_2}&=& -f_1f_2\frac{\varepsilon}{2}
   +f_1 f_2 (f_1 + f_2 -3)\frac{\varepsilon^{2}}{8}
- f_1  f_2  (f_1 + f_2 - 3)  \Big [3 \zeta(3)+ f_1
\\ \nonumber &&
+ f_2 - 3  \Big ]\frac{\varepsilon^{3}}{16} - f_1\, f_2\,   \Big [
70\, \zeta (5 ) - 54\, \zeta (4 ) - 56\, \zeta (3 ) - 132\, f_1
\\ \nonumber &&
- 132\, f_2 + 60\, \zeta (3 )\, f_1 + 60\, \zeta (3 )\, f_2 + 18\,
\zeta (4 )\, f_1 + 18\, \zeta (4 )\, f_2
\\ \nonumber &&
+ 30\, \zeta (5 )\, f_1 + 30\, \zeta (5 )\, f_2 + 44\, f_1^2 - 5\,
f_1^3 + 44\, f_2^2 - 5\, f_2^3 + 96\, f_1\, f_2
\\ \nonumber &&
- 8\, \zeta (3 )\, f_1^2 - 8\, \zeta (3 )\, f_2^2 - 20\, \zeta (5
)\, f_1^2 - 20\, \zeta (5 )\, f_2^2 - 42\, \zeta (3 )\, f_1\, f_2
\\ \label{8} &&
- 30\, \zeta (5 )\, f_1\, f_2 - 16\, f_1\, f_2^2 - 16\, f_1^2\,
f_2 + 125  \Big ]\frac{\varepsilon^{4}}{128},
\end{eqnarray}
\begin{eqnarray}\nonumber
\nonumber \eta^U_{f_1f_2}&=&  f_1 (1-f_1 - 3 f_2)
\frac{\varepsilon}{8} +f_1 (8 f_1^2 - 91 f_2 - 33  f_1 + 18  f_2^2
+ 42  f_1  f_2
\\ \nonumber &&
+ 42\, \zeta(3)\, f_1\, f_2 + 25)\frac{\varepsilon^{2}}{256} - f_1
 \Big [ 712  \zeta(3) + 969  f_1  + 2463 f_2
\\ \nonumber &&
- 936  \zeta(3)  f_1
- 2652 \zeta(3)  f_2 - 456  f_1^2 + 64 f_1^3
- 1050  f_2^2 + 108  f_2^3
\\ \nonumber &&
- 2290  f_1 f_2 + 224 \zeta(3) f_1^2 + 540  \zeta(3)  f_2^2 + 1188
\zeta(3) f_1 f_2 + 504  f_1 f_2^2
\\ \nonumber &&
 +
492  f_1^2  f_2 - 577  \Big ]\frac{\varepsilon^{3}}{4096} +f_1\,
 \Big [ 99200\, \zeta (5 ) - 34176\, \zeta (4 ) - 77136\, \zeta (3 )
\\ \nonumber &&
- 136137\, f_1 - 321075\, f_2 + 128528\, \zeta (3 )\, f_1 +
286200\, \zeta (3 )\, f_2
\\ \nonumber &&
+ 44928\, \zeta (4 )\, f_1 + 127296\, \zeta (4 )\, f_2 - 99840\,
\zeta (5 )\, f_1 - 311280\, \zeta (5 )\, f_2
\\ \nonumber &&
+ 94376\, f_1^2 - 26752\, f_1^3 + 216098\, f_2^2 + 2688\, f_1^4 -
46872\, f_2^3
\\ \nonumber &&
+ 3240\, f_2^4 + 456730\, f_1\, f_2 - 59712\, \zeta (3 )\, f_1^2 +
8320\, \zeta (3 )\, f_1^3
\\ \nonumber &&
- 122904\, \zeta (3 )\, f_2^2 - 10752\, \zeta (4 )\, f_1^2 +
10368\, \zeta (3 )\, f_2^3 - 25920\, \zeta (4 )\, f_2^2
\\ \nonumber &&
- 8960\, \zeta (5 )\, f_1^2 + 9600\, \zeta (5 )\, f_1^3 - 36720\,
\zeta (5 )\, f_2^2 + 21600\, \zeta (5 )\, f_2^3
\\ \nonumber &&
- 282504\, \zeta (3 )\, f_1\, f_2 - 57024\, \zeta (4 )\, f_1\, f_2
- 47280\, \zeta (5 )\, f_1\, f_2
\\ \nonumber &&
- 212496\, f_1\, f_2^2 - 199768\, f_1^2\, f_2 + 24192\, f_1\,
f_2^3 + 27624\, f_1^3\, f_2
\\ \nonumber &&
+ 72144\, \zeta (3 )\, f_1\, f_2^2 + 59328\, \zeta (3 )\, f_1^2\,
f_2 + 79920\, \zeta (5 )\, f_1\, f_2^2
\\ \label{9} &&
+ 80160\, \zeta (5 )\, f_1^2\, f_2 + 47808\, f_1^2\, f_2^2 + 65825
 \Big ] \frac{\varepsilon^{4}}{262144},
\end{eqnarray}
\begin{eqnarray}
\nonumber \eta^{{\rm MAW}}_f &=&- f (f - 1)\frac{\varepsilon}{4} +
f (f - 1) (2 f - 5) \frac{\varepsilon^{2}}{16} - f (f - 1) \Big [
8 \zeta(3)  f - 20  f
\\ \nonumber &&
- 19  \zeta(3) + 4  f^2 + 25  \Big ] \frac{\varepsilon^{3}}{32} +
f\,  (f - 1 )\,   \Big [ 782\, f + 222\, \zeta (3 )
\\ \nonumber &&
+ 114\, \zeta (4 ) + 120\, \zeta (5 ) - 230\, f\, \zeta (3 ) -
48\, f\, \zeta (4 ) - 310\, f\, \zeta (5 )
\\ \label{10} &&
- 314\, f^2 + 42\, f^3 + 58\, f^2\, \zeta (3 ) + 110\, f^2\, \zeta
(5 ) - 647  \Big ] \frac{\varepsilon^{4}}{256}.
\end{eqnarray}
In (\ref{8})-(\ref{10}), $\zeta(x)$  is the Riemann zeta-function.
The above expressions recover the $\varepsilon^3$ results of Ref.
\cite{Ferbercopol} and describe the scaling of star-like
combinations of different species: two sets of RWs B, B$^\prime$
(\ref{8}); $f_1$ SAWs and $f_2$ RWs, A and B (\ref{9}); $f$ MAWs,
C (\ref{10}). The case of the homogeneous star exponent is
included via $\eta^S_{f_A}=\eta^U_{f_A0}$.

%%%%%%%%%%%%%%%%%%%%%%%%%%%%
%SECTION IV
%%%%%%%%%%%%%%%%%%%%%%%%%%%%

\section{Conclusions and outlook}
\label{IV}

The expansions (\ref{8})-(\ref{10}) are the starting point for the
construction of analytic expressions for a number of physical
quantities describing various phenomena (we mention some of them
in the section \ref{I}): the exponents govern the short distance
interaction of star polymers in colloidal solutions
\cite{note1,Schulte-2003a} as well as the reaction rate of
diffusion controlled reactions with traps or reaction sites
attached to polymer chains or to a star polymer
\cite{note1,Cates87}. One more recent application concern scaling
of distribution of denaturated loops and unzipped end segments for
a model of DNA denaturation \cite{Baiesi02}. Moreover, due to the
special convexity properties of the spectrum of copolymer star
exponents \cite{FerberMF,Duplantier91} they can be re-written in
the multifractal formalism, in terms of a spectral function and
H\"older exponents \cite{Cates87,Duplantier99,FerberMF}. Such a
program is well outside the scope of this Letter and is the
subject of separate publications
\cite{Schulte-2003,Schulte-2003a}. We also hope that our study
will motivate the analysis of these and similar problems by other
tools.

\begin{figure}[htbp]
\includegraphics[width=13.5cm]{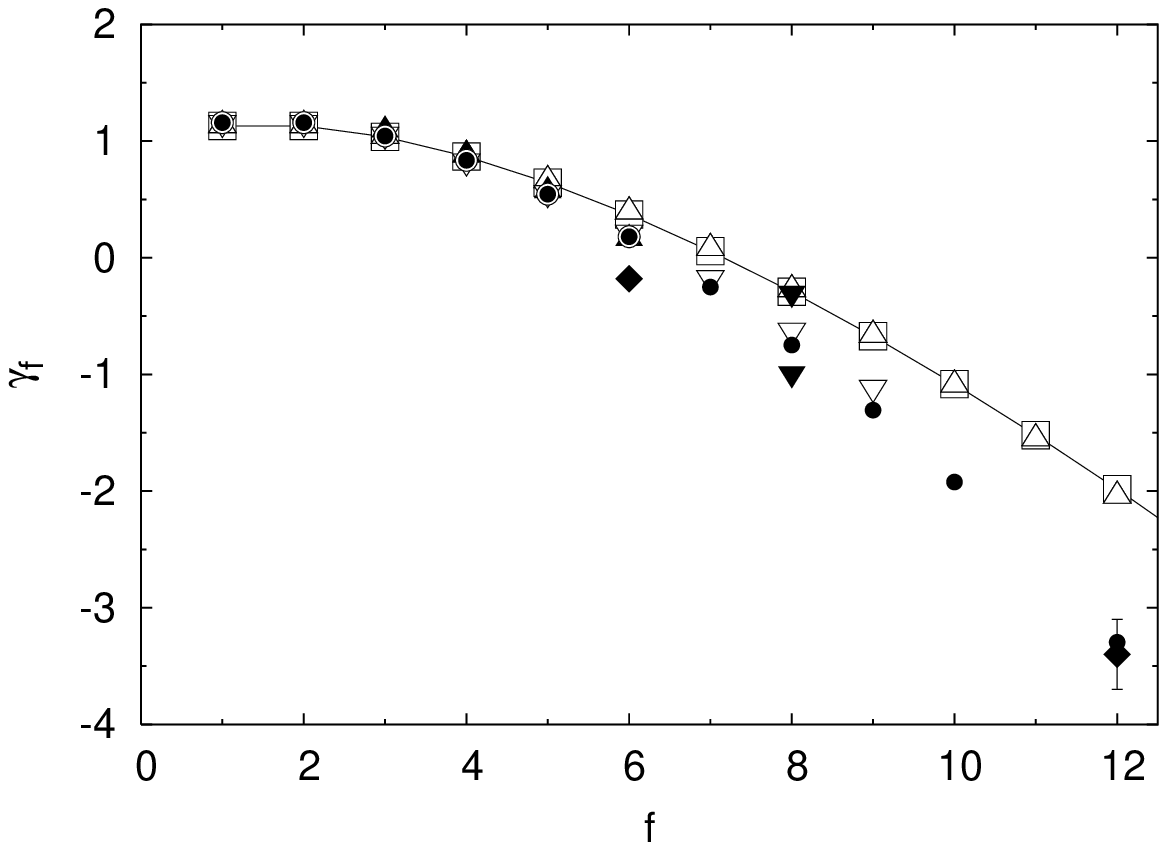}
\caption{ \label{fig1} Homogeneous star exponent $\gamma_f$ at
$d=3$. Open squares: forth order ($\varepsilon^4$), open triangles
up:  third order ($\varepsilon^3$), open triangles down
\protect\cite{Ferber-1996b}: third order RG calculated at fixed
dimension ($d=3$). Results of MC simulations are shown by filled
symbols, triangles up: \protect\cite{Batoulis89}, triangles down:
\protect\cite{Barrett87}, diamonds: \cite{Ohno94}, discs:
\cite{Grassberger03}.}
\end{figure}

As a final note, we want to emphasize another particular feature
of the spectra of exponents (\ref{8})-(\ref{10}) distinguishing
them from the usual $\varepsilon$-expansions obtained within
$\phi^4$ theory. The field theoretic perturbative expansions are
known \cite{RGbooks} to be asymptotic at best and one should apply
a resummation procedure to calculate the numerical estimates from
these series. Such a resummation allows, in particular, to restore
the convergence of the series in $\varepsilon$. However, the
series (\ref{8})-(\ref{10}) posses also a strong $f$-dependence.
The standard resummation technique, leading to high-precision
estimates in the field-theoretical scheme will give similar
accuracy for low values of $f$ only. To show this and to give an
idea of the numerical estimations, based on e.g. Borel resummation
refined by a conformal mapping \cite{RGbooks}, we display one
particular result in Fig. \ref{fig1}. Here, we plot our resummed
$\varepsilon^4$ results for the star exponent
$\gamma_f=1+\nu(\eta^U_{f0}-f\eta^U_{20})$ \cite{Ferbercopol} with
the Flory exponent $\nu=0.588$ in $d=3$. This exponent governs the
scaling of the homogeneous star partition function ${\mathcal
Z}^{(*f)}$ with respect to the number of monomers $N$ in the
chain: ${\mathcal Z}^{(*f)}\sim N^{\gamma_f-1}$. We compare the
resummed $\varepsilon^4$ data with the results of MC simulations
\cite{Batoulis89,Barrett87,Ohno94,Grassberger03} and with the
resummed three-loop pseudo-$\varepsilon$ expansion obtained within
the massive RG scheme at fixed $d=3$ \cite{Ferber-1996b}. One
observes a nice correlation within successive perturbative
approximations and MC simulations for small $f$ and an expected
growing discrepancy when increasing the number of chains. A recent
MC simulation \cite{Grassberger03} has determined values for the
exponents with high precision. One sees from the figure \ref{fig1}
that the usual resummation does not meet these values for high
$f$. We believe that a more sophisticated technique should also
include the known large-$f$ behaviour.

%%%%%%%%%%%%%%%%%%%%%%%%%%%%
%SECTION V
%%%%%%%%%%%%%%%%%%%%%%%%%%%%

\section*{Acknowledgements}

This work was supported in part by the Deutsche
Forschungsgemeinschaft and by the Austrian Fonds zur F\"orderung
der wissenschaftlichen Forschung, project No.16574-PHY. C.v.F. and
Yu.H. acknowledge the kind hospitality and friendly attention of
Reinhard Folk in Linz, where this work was completed.

%\bibliographystyle{elsart-num}
%\bibliography{cv_publist,ferber,star4}

\end{document}